\documentstyle[12pt,epsfig]{article}
\begin{document}
\noindent
\begin{center}
{\Large {\bf  Gravitational Coupling, Dynamical Changes of Units and The Cosmological Constant Problem}}\\
\vspace{2cm}
 ${\bf Yousef~Bisabr}$\footnote{e-mail:~y-bisabr@sru.ac.ir.}\\
\vspace{0.5cm} {\small{Department of Physics, Shahid Rajaee Teacher
Training University,
Lavizan, Tehran 16788, Iran}}\\
\end{center}
\vspace{1cm}
\begin{abstract}
A spacetime interval connecting two neighbouring points can be measured in different unit systems. For instance, it can be measured in atomic unit defined in terms of fundamental constants existing in quantum theories. It is also possible to use a gravitational unit which is defined by using of properties of macroscopic objects. These two unit systems are usually regarded as indistinguishable up to a constant conversion factor. Here we consider the possibility that these two units are related by an epoch-dependent conversion factor. This is a dynamical changes of units.
Regarding a conformal transformation as a local unit transformation, we use a gravitational model in which the gravitational and the matter sectors are given in different conformal frames (or unit systems). It is relevant to the cosmological constant problem, namely the huge discrepancy between the estimated and the observational values of the cosmological constant in particle physics and cosmology, respectively. We argue that the problem arises when one ignores evolution of the conversion factor relating the two units during expansion of the Universe.  Connection of the model with violation of equivalence principle and possible variation of fundamental constants are also discussed.
\end{abstract}
~~~~~~~~~~~~~~~~~~~~~~~~~~~~~~~~~~~~~~~~~~~~~~~~~~~~~~~~~~~~~~~~~~~~~~~~~~~~~~~~~~~~~~~~~~~~~~~~~~~~~
\section{Introduction}
In a gravitational theory, there is an implicit assumption concerning coupling of matter systems with gravity: all types of matter should couple with a unique metric that describes the background
geometry. This universality of gravitational coupling is supported by equivalence principle (EP) and
was considered by Einstein as a basis to formulate the theory of General Relativity.  This principle is a heuristic generalization of the experimental
fact that all neutral bodies, independent of the details of compositions and internal structures,  seem to fall with the same acceleration in an external
gravitational field.  In a particle physics language, the EP is the assumption that all Standard Model (SM) fields, in spite of wide variety of physical properties, couple
in the same manner with gravity. The EP has many observable consequences and has been verified experimentally many
times since Galileo \cite{1}.  Despite these experimental verifications, some authors seem to be
unsatisfied with theoretical status of the principle and thus consider a possibility of its violation at some
levels. For instance, there is a strong debate about validity of the EP in quantum domains \cite{2}. In particular, string theory suggests
the existence of some scalar fields (dilaton or moduli) whose interactions
with matter induce violation of the EP \cite{3}. Such a violation is not only expected in quantum gravity
domain but it is also predicted in a semiclassical regime\footnote{In this framework the matter is described by quantum field theory while
the gravitational field itself is regarded as a classical object.}. To clarify the latter point consider the Galilean statement
 of EP, namely that all freely falling bodies follow the same trajectories in a gravitational field with the same initial conditions. When
 matter is supposed to behave quantum mechanically, particles do no longer follow a well-defined classical trajectory (due to the uncertainty principle) and
thus the EP is expected not to be satisfied.\\
Apart from possible violation of the EP in quantum regimes, there is also a possibility of such a violation even with a purely classical
treatment. We note that although all EP tests
have been done in a limited time interval (about four hundred years since Galileo), the results
are extrapolated to entire age of the Universe. It is, however, quite possible that the EP has been
violated in some stages during evolution of the Universe. For instance, consider a gravitational massless scalar field which interacts with
matter \cite{4} and thus induces violation of the EP. In general, the strength of such a coupling
should not be taken as a constant and it may vary during expansion of the Universe. If it weakens then it disappears from dynamics of the Universe
at sufficiently late times. In this case, the EP and the universality of gravitational coupling is a late-time manifestation
of gravitational theories.\\
Another possibility is that the scalar field is a part of a screening mechanism and hides itself in experiments.
For instance, if it is taken to be a chameleon field \cite{5}, then its interaction with matter can not be detected in experiments. In this case, the
chameleon is heavy enough in the environment of the laboratory tests so that the local gravity
constraints suppressed. Meanwhile, it can be light enough in the low-density cosmological
environment to have observable effects at large scale.\\
These arguments prompt us to open our mind about matter gravitational coupling in a gravitational model. We remark that investigation of the possibility that matter does not couple to gravity as suggested by EP, opens new windows for addressing some challenging problems in cosmology. The present work deals with presentation of a more general framework for gravitational coupling of matter systems. A particular application of this framework to the cosmological constant problem can be found in \cite{bis}. The article is organized as follows : In section 2, we consider a conformal transformation as a local unit transformation, namely a transformation of units with a spacetime dependent conversion factor. We explore the conjecture that ratio of a spacetime interval in gravitational and atomic units, may have not been the same throughout the evolution of the Universe. We argue that this scheme is extremely relevant to the cosmological constant problem. In section 3, we will use this scheme to construct a gravitational model in which the gravitational and matter sectors belong to two different conformal frames (CFs) (or unit systems). This differs from the case that only the metric of the two sectors are related by a conformal transformation (CT) \cite{g1} \cite{g2}. Here the whole matter sector, including the metric and all matter fields as dimensional quantities, lives in a different CF. The conformal factor appears then as a dynamical scalar field. By finding exact solutions of the corresponding field equations in a cosmological context, we show that evolution of this scalar field during expansion of the Universe leads to reduction of a large cosmological constant. In section 4, we draw our conclusions.
\section{Conformal transformation as a unit transformation}
\subsection{Conformal transformations}
For $(M, g_{\mu\nu})$ as a manifold endowed with a metric, a line element is
given by $ds^2=g_{\mu\nu}dx^{\mu}dx^{\nu}$. If the spacetime metric $g_{\mu\nu}$ is regarded
as carrying dimensions while coordinates are taken to be dimensionless, a CT is
defined by \cite{con}
\begin{equation}
\bar{g}_{\mu\nu} =e^{-2\sigma}  g_{\mu\nu} \label{1-1}\end{equation}
where
$\sigma=\sigma(x^{\mu})$ is a positive, smooth and dimensionless function. The line element transforms as
\begin{equation}
ds\rightarrow d\bar{s}=e^{-\sigma}ds
\label{1-1a}\end{equation}
Therefore a CT
is equivalent to a change of spacetime intervals and may be interpreted as a unit transformation. Contrary to
a global unit transformation in which the conversion factor is a constant\footnote{For instance, a change of mks
unit system to cgs is a global unit transformation.}, a CT is actually a local
unit transformation \cite{bek} \cite{dicke}. In a CT all dimensional quantities should be changed
according to their dimensions. For instance, if $\chi$ is a SM field then
\begin{equation}
\bar{\chi} = e^{\gamma\sigma} \chi \label{1-2}\end{equation}
where the parameter $\gamma$ is a weight determining that $\chi$ corresponds to which of the
SM fields.  For instance, for scalar fields $\gamma=1$
and for Dirac fields $\gamma$=3/2 \cite{bek}.
\subsection{Dynamical changes of units}
There are two distinct unit systems which are defined in terms of properties of two fundamental theories that revolutionized physics in the twentieth
 century: Quantum Mechanics and General Theory of Relativity. The atomic unit system is made of constants such as charges ($e$) and masses ($m$) of elementary particles,
$\hbar$ and $c$. On the other hand, the gravitational unit system is made of constants such as $G$, masses ($M$)
and sizes ($R$) of macroscopic objects. Each of these systems contains a complete set of constants
in the sense that one can use the elements of each set to construct units of time, length and mass without need to invoke elements of the other set. It is a general assumption that transformation from one unit to the other is performed by a constant conversion factor. This implies
that each of these unit systems is a constant multiple of the other. For instance, if $dS_{G}$ and $dS_{A}$ are spacetime
intervals in gravitational and atomic units, respectively, then $dS_{G}=\beta~ dS_A$ which $\beta$ is a constant.
The conversion factor $\beta$ itself is given in terms of fundamental
constants $e$, $\hbar$, $c$, $G$ and so on. Evidently, the assumption that $\beta$ is a constant means constancy of these fundamental physical quantities. In this case, the use of each unit system is a matter of convenience and is devoid of
any dynamical meaning. It means that one may arbitrarily use the atomic unit system
to describe the evolution of the Universe or the gravitational unit system to
describe dynamical properties of an elementary particle. \\
There are some arguments against the constancy of $\beta$ started by Dirac's Large Number Hypothesis \cite{d1}. He supposed that the spacetime interval $dS_{G}$ defined in a gravitational theory is not the same as $dS_{A}$ measured
by an atomic apparatus. By taking the ratio of the two line elements to vary with the epoch, he formulated a $G$-varying theory and reformulated Weyle's geometry \cite{d2}. Later, Canuto et al constructed
a scale-covariant gravitational theory and studied some of its astrophysical consequences \cite{can}. In this bimetric viewpoint, the quantities $e$, $m$ and $\hbar$ are constants with respect to atomic units while $G$, $M$ and
$R$ vary. On the other hand, in gravitational units the reverse is true. It should be remarked that the dynamical distinction\footnote{This means that the two unit systems are connected by a non-constant conversion factor.} of theses unit systems is closely related to the arguments concerning variations of fundamental physical constants. Since $\beta$ is given in terms of some physical constants existing in quantum and gravitational theories, any indication of variations of these constants implies changes of $\beta$ with the epoch. Even though there is not still a verified observation indicating such a variation, any future indication of such variations emphasizes that the gravitational and the atomic unit systems must be considered as dynamically distinct.\\
Let us investigate the consequences of the following assumption
\begin{equation}
dS_G=\beta(t)~ dS_A
\label{1-8}\end{equation}
where the scale function $\beta(t)$ is now a function of the epoch. Comparing this with (\ref{1-1a}) reveals that this is actually a CT with conformal factor $\beta^{-1}$\footnote{In the following, the atomic unit system is denoted by barred quantities.}. Since the ratio of a particular
spacetime interval in gravitational and atomic units enlarges in an expanding Universe, the expression (\ref{1-8}) suggests that $\beta(t)$ should be an increasing function of time. This gives a dynamical meaning to changes of unit systems.
We emphasize that this leads to an ambiguity when one decides to compare a physical quantity in
a gravitational theory (or cosmology) with its corresponding value in quantum (or elementary particle) physics. For instance, let us consider the cosmological constant. The theoretical estimations
on the value of this quantity in particle physics are made in the atomic units which differs from the unit that the cosmological observations are
made. The cosmological constant problem arises
when one \emph{ignores} the dynamics of $\beta(t)$. The huge discrepancy between observations and theoretical estimates is the result of taking the two  unit systems indistinguishable up to a constant conversion factor. However, the discrepancy may be removed when the unit transformation is regarded as a dynamical process rather than a naive multiplication by a constant factor \cite{bis}.

\section{Anomalous gravitational coupling }
We consider a gravitational model whose matter and gravitational parts belong to
different CFs. The total action of such a model can be written as\footnote{We take the gravitational part to be described by the Einstein-Hilbert term but it may be generally replaced by a modified gravitational theory.}
\begin{equation}
S=\frac{1}{2\kappa} \int d^{4}x \sqrt{-g} R -\int d^{4}x
\sqrt{-\bar{g}} L(\bar{g}_{\mu\nu},\bar{\chi})
\label{1}\end{equation}
 where $\kappa=8\pi M_p^{-2}$ with $M_p$ being the Planck mass. The first term is the Einstein-Hilbert
 action characterizing the background geometry and the second term is the action of all matter fields
  which is collectively denoted by $\bar{\chi}$. Note that the whole of the matter part, which contains the metric and all possible SM fields, belongs to a CF which is different from that of the gravitational part. When $\sigma=const.$, the CF corresponding to the matter system is related to the gravitational frame by a global unit transformation and the gravitational coupling of matter encoded in (\ref{1}) coincides with the standard picture. This means that in that case the gravitational and the matter sectors are given in units that are constant multiple of each other. On the other hand, when $\sigma$ varies one expects that the EP does not hold. We emphasize that this does not immediately mean that such a model contradict with today's experiments.  Experimental consistency crucially depends on the dynamics of $\sigma$ . For instance, suppose that the variation of $\sigma$ becomes relevant at early times and that this variation diminishes so that $\sigma$ takes a constant configuration at sufficiently late times. In this case, the late-time behavior of the model is the same as the standard picture, namely that the gravitational and the atomic units are related by a constant conversion factor and all matter components in (\ref{1}) normally couple to the gravitational part. In such a case, one can say that the EP is only a late-time manifestation of the model\footnote{ The EP violation at early times may cause successful parts of the hot big bang model, such as
the theories of primordial Nucleosynthesis and structure formation, to be faced with serious difficulties. To avoid such difficulties one may push the violation at some earlier times. e.g., in the inflationary era \cite{arx}.}(\ref{1}).
 \\
 We would like to apply the action (\ref{1}) to a Universe in which there is a scalar field component whose energy density
 dominates over all other energy densities. Such a Universe may be taken as a model for our Universe
 in the inflaton-dominated (or inflationary) regime at early times. It may also be corresponded to late-time
 evolution of the Universe dominated by a quintessence field.
In this case, we have
\begin{equation}
S=\frac{1}{2\kappa} \int d^{4}x \sqrt{-g} R -\int d^{4}x
\sqrt{-\bar{g}}
~\{\frac{1}{2}\bar{g}^{\mu\nu}\nabla_{\mu}\bar{\phi}\nabla_{\nu}\bar{\phi}+V(\bar{\phi})\}
\label{1-4}\end{equation}
where $\bar{\phi}$ is a scalar field with potential $V(\bar{\phi})$ written in atomic units. In terms of the background
variables ($g_{\mu\nu}$, $\phi$), this action takes the form
\begin{equation}
S=\frac{1}{2} \int d^{4}x \sqrt{-g}~ \{\frac{1}{\kappa}R-g^{\mu\nu}
\nabla_{\mu} \phi \nabla_{\nu}\phi-\phi g^{\mu\nu} \nabla_{\mu} \phi
\nabla_{\nu}\sigma-\phi^2 g^{\mu\nu} \nabla_{\mu}
\sigma \nabla_{\nu}\sigma-V(e^{\sigma}\phi)e^{-4\sigma}\}
\label{1-6}\end{equation}
It is important to note that the conformal factor $\sigma$ appears as a dynamical field.
The above
expression is the action functional of two
dynamical scalar fields with a mixed kinetic term \cite{k1}
\cite{k2}. Such a system is used in Literature to formulate assisted
quintessence \cite{quin} and to improve dark energy models \cite{k2}
\cite{da}. \\
Let us consider a region that $\phi$ is a sufficiently slowly varying field\footnote{ At early inflaton-dominated (or inflationary) regime, this
is the slow-roll approximation \cite{sr}.} ($\phi\approx const.$). In general, the equation of state parameter of $\phi$ is
\begin{equation}
\omega_{\phi}=\frac{\frac{1}{2}\dot{\phi}^2-V(\phi)}{\frac{1}{2}\dot{\phi}^2+V(\phi)}
\end{equation}
In a slowly varying region, $\omega_{\phi}\approx -1$ and $V(\phi)$ mimics a cosmological term. In this approximation, (\ref{1-6}) reduces to
\begin{equation}
S=\frac{1}{2} \int d^{4}x \sqrt{-g}~ \{\frac{1}{\kappa}R-\phi^2
g^{\mu\nu}
\nabla_{\mu} \sigma \nabla_{\nu}\sigma-V(e^{\sigma}\phi)e^{-4\sigma}\} \label{1-7}\end{equation}
Variation of (\ref{1-7}) with respect to $g_{\mu\nu}$ and $\sigma$ gives the corresponding field
equations. In a spatially flat Friedmann-Robertson-Walker
cosmology, they take the form
\begin{equation}
3 H^2=\frac{1}{2}\kappa
(\phi^2\dot{\sigma}^2+V e^{-4\sigma})
\label{1-9}\end{equation}
\begin{equation}
\ddot{\sigma}+3H \dot{\sigma}+
\frac{1}{2\phi^2}\frac{d}{d\sigma} (Ve^{-4\sigma})=0
\label{1-10}\end{equation} where $H=\frac{\dot{a}}{a}$ with $a$ being the scale factor and $\sigma$ is taken to be only a
function of time in a homogenous and isotropic Universe. In these equations, the exponential coefficient in front of the potential function
is relevant in the cosmological constant problem. If $\sigma$ is an increasing function of time, then $e^{-4\sigma}$ acts as a damping coefficient which induces reduction of a large cosmological term during expansion of the Universe.\\
As an illustration, let us
consider the equations (\ref{1-9}) and (\ref{1-10}) for a monomial potential
\begin{equation}
V(\bar{\phi})= \lambda M_p^4(\frac{\bar{\phi}}{M_p})^n
\label{pot}\end{equation}
 with $n$ and $\lambda$ being constants. In this case, there is an exact solution \cite{arx}
\begin{equation}
a(t)=a_0 t^p \label{n}\end{equation}
\begin{equation}
\sigma(t)=\sigma_0+\frac{2}{(4-n)}\ln (\frac{t}{t_0} )
\label{1-11}\end{equation} with
\begin{equation}
p=\frac{16\pi \alpha^2}{(n-4)^2}
\label{p}\end{equation}
\begin{equation}
M_p t_0=[\frac{4(3p-1)}{\lambda \alpha^{n-2}(n-4)^2 e^{(n-4)\sigma_0}}]^{\frac{1}{2}}
\label{2-11}\end{equation}
where $\alpha\equiv\phi/M_p$ and $\sigma_0$ is a dimensionless constant. This solution indicates that the Universe experiences an accelerating expansion for $\alpha>\frac {|n-4|}{4\sqrt{\pi}}$. Defining $\Lambda_{eff}\equiv \frac{1}{2}\kappa V(\bar{\phi})e^{-4\sigma}= 4\pi \lambda M_p^2 \alpha^n e^{(n-4)\sigma}$ and combining the latter with (\ref{1-11}) reveals that $\Lambda_{eff}$ actually decays
and scales as $\Lambda_{eff}\sim t^{-2}$. Thus anomalous gravitational coupling of $\phi$, whose its potential acts as a large cosmological term, is responsible for both an accelerating expansion of the Universe and damping of $\Lambda_{eff}$ during the expansion.\\
As the last point, we remark that in our strategy all matter species couple anomalously with gravity as indicated by (\ref{1}). Details of such a  coupling is given by the field $\sigma$ which its behavior is illustrated in the following figures. The figures show that after a sharp variation, $\sigma(t)$ immediately takes a constant configuration.
\begin{figure}[ht]
\begin{center}
\includegraphics[width=0.45\linewidth]{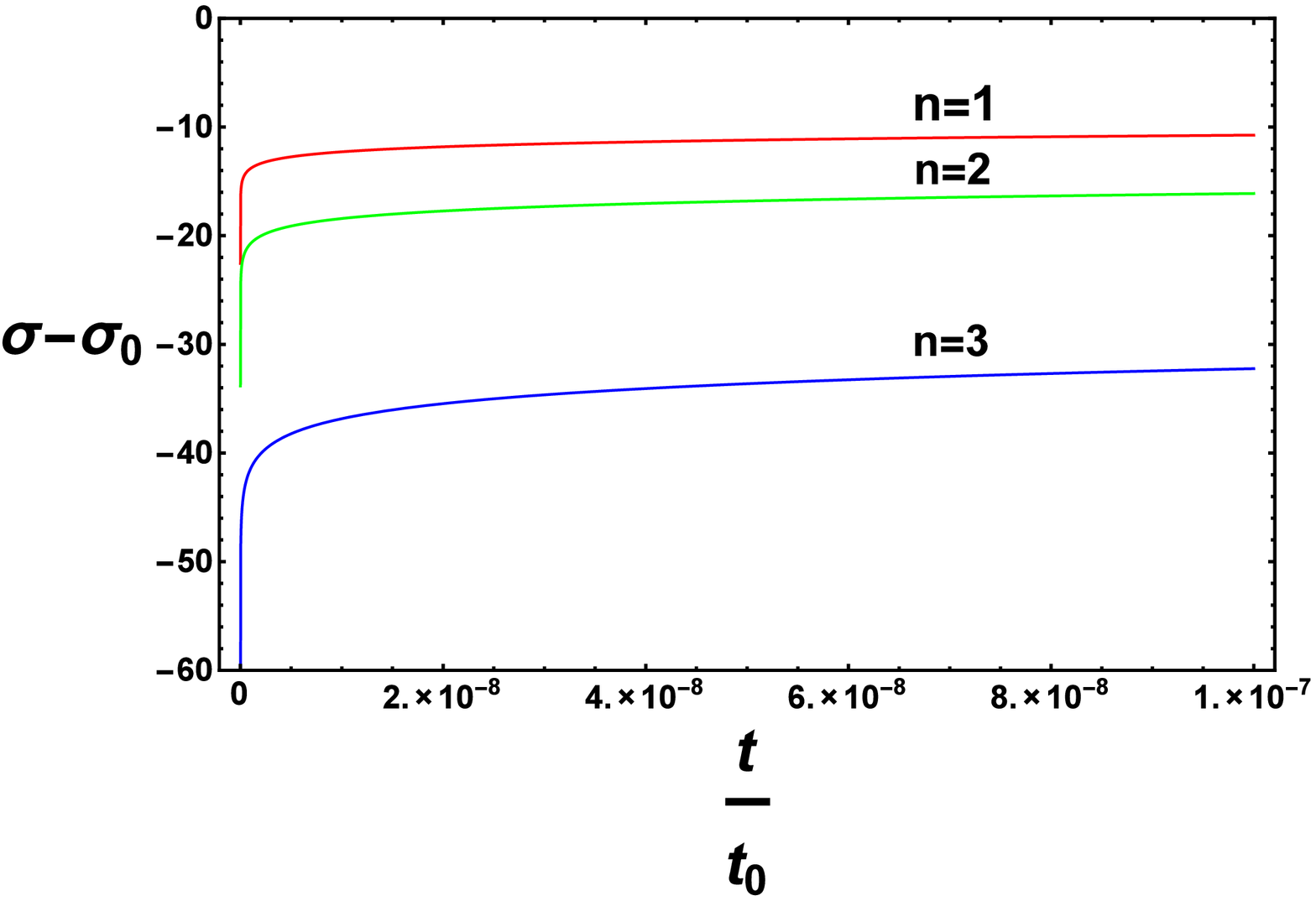}
\includegraphics[width=0.45\linewidth]{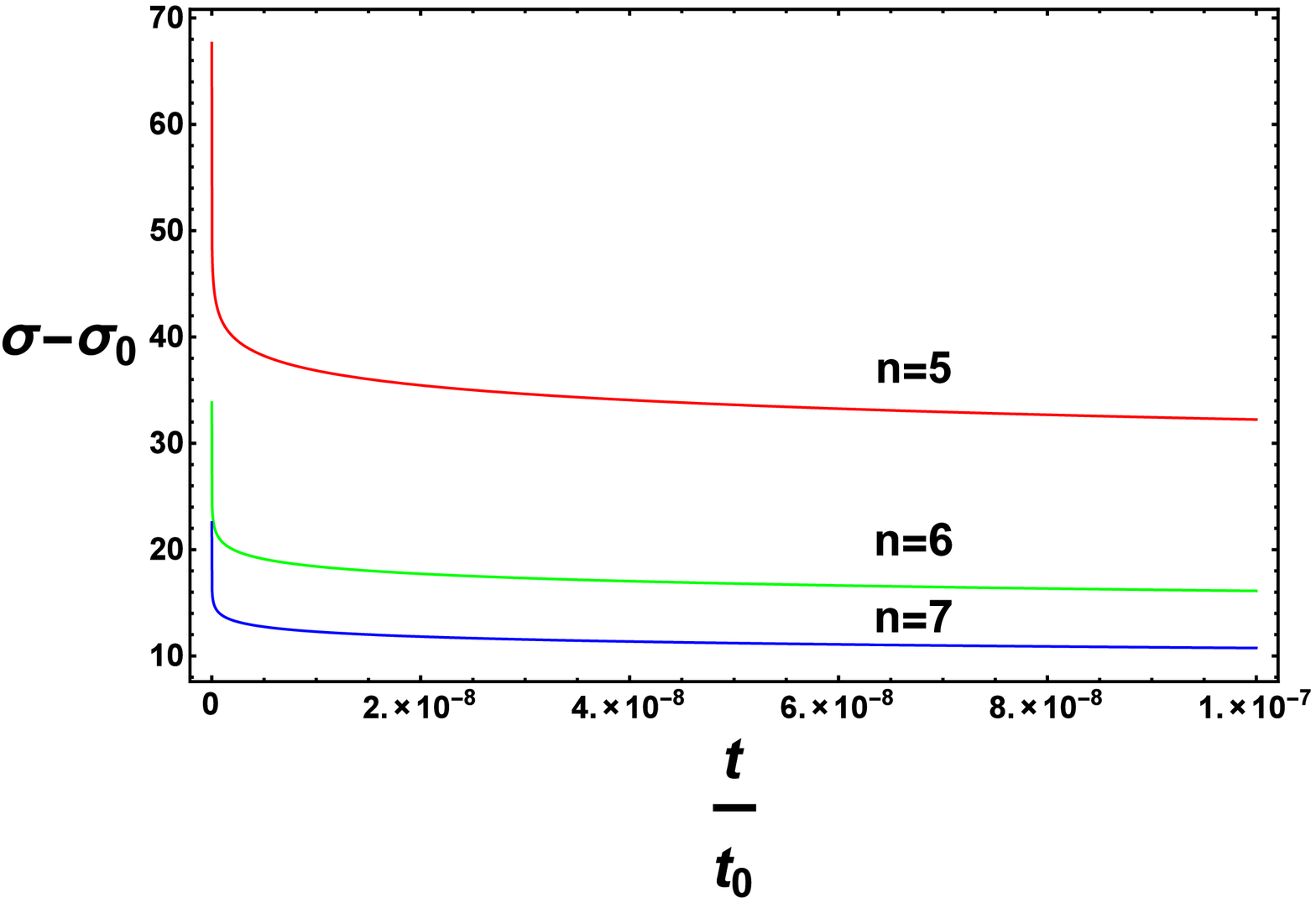}
\end{center}
\end{figure}
Then the two metrics $g_{\mu\nu}$ and $\bar{g}_{\mu\nu}$ become dynamically equivalent. This dynamic equivalence implies that the gravitational coupling of matter become normal and the action (\ref{1}) is reduced to
\begin{equation}
S=\frac{1}{2\kappa} \int d^{4}x \sqrt{-g} R -\int d^{4}x
\sqrt{-g} L(g_{\mu\nu},\chi)
\label{1a}\end{equation}
which is in accord with EP and the standard picture of matter coupling in gravitational theories.
\section{Discussion}
We have studied some consequences of a gravitational model whose vacuum and matter sectors are defined in terms of gravitational and atomic units, respectively. Contrary to the usual viewpoint, these two unit systems are connected by an epoch dependent conversion factor.
 This is related to the basic idea of the Dirac's bimetric proposal \cite{d2} \cite{can} and differs necessarily from other bimetric theories which try to model possible speed of light variations \cite{vsl} or massive gravity \cite{bi}. \\
We argue that regarding transformation of these units as dynamical is extremely relevant to the cosmological constant problem. In our analysis, this problem may be alleviated if one takes into account evolution of the field $\sigma$ (the conversion factor) during expansion of the Universe.
This idea is closely related to the possibility that some fundamental physical constants have varied during some stages of evolution of the Universe. Therefore any future indication of such variations will be a direct justification of regarding the two unit systems as dynamically distinct. In this sense, the cosmological constant problem is also connected to whether or not some fundamental physical constants have been changed throughout the evolution of the Universe. \\

\end{document}